\documentclass{eptcs}
\providecommand{\event}{ACL2 2011} 
\usepackage{breakurl}             

\providecommand{\urlalt}[2]{\href{#1}{#2}}
\providecommand{\doi}[1]{doi:\urlalt{http://dx.doi.org/#1}{#1}}

\usepackage{hyperref}
\usepackage{acl2}
\usepackage{amsmath,amssymb,enumerate}
\usepackage{graphicx}
\usepackage{listings}
\usepackage{url} 
\title{How Can I Do That with ACL2? \\ Recent Enhancements to ACL2}

\author{
    Matt Kaufmann\\
    \institute{Dept. of Computer Science,\\
               University of Texas at Austin}
    \email{kaufmann@cs.utexas.edu}
    \and
    J Strother Moore\\
    \institute{Dept. of Computer Science,\\
               University of Texas at Austin}
    \email{moore@cs.utexas.edu}
    }

\newcommand{\question}[1]{\vspace{0.1 in}\noindent {\bf \underline{Question}}: {\em #1}\medskip}

\def\titlerunning{Recent Enhancements to ACL2}
\def\authorrunning{Matt Kaufmann and J Strother Moore}

\begin{document}
\maketitle
\lstset{language=Lisp}


\begin{abstract}

The last several years have seen major enhancements to ACL2
functionality, largely driven by requests from its user community,
including utilities now in common use such as {\tt make-event}, {\tt
  mbe}, and trust tags.  In this paper we provide user-level summaries
of some ACL2 enhancements introduced after the release of Version 3.5
(in May, 2009, at about the time of the 2009 ACL2 workshop) up through
the release of Version 4.3 in July, 2011, roughly a couple of years later.  Many of
these features are not particularly well known yet, but most ACL2
users could take advantage of at least some of them.  Some of the
changes could affect existing proof efforts, such as a change that
treats pairs of functions such as {\tt member} and {\tt member-equal}
as the same function.

\end{abstract}

\section{Introduction}

This paper discusses ACL2 enhancements made in ACL2 Versions 3.6 through 4.3,
that is, during the 2+ years that have passed since Version 3.5 was
released around the time of the preceding (2009) ACL2
workshop.  These enhancements primarily concern programming, proof control,
and system infrastructure --- as opposed to improved proof procedures,
sophisticated logical extensions, and theoretical studies.  Readers from
outside the ACL2 community --- should there be any! --- may find this pragmatic
stance surprising.  But ACL2's total integration of programming, proof, and
system issues is one of the reasons ACL2 finds industrial application and we
believe that the research issues raised by our commitment to integration are
at least as important as more conventional theorem proving work.

Even though ACL2 is typically
modified in response to user requests, still we suspect that most
recent enhancements are unknown to most ACL2 users.  Perhaps that is
because the release notes for the above versions of ACL2 list about
300 improvements, hence serving as a large, rather flat and complete
reference document that one may prefer not to read carefully.  Our
goal here is to raise awareness of the most important of these
enhancements.

Our focus is on the user level, both here and in the release notes.
This paper thus includes many examples.  We do not claim that
this paper covers every interesting enhancement.  A more complete summary
of changes can be found in the release notes and various documentation
topics.  Indeed, as our goal is to bring awareness of recent ACL2
changes to the community, even the topics that we do cover in this
paper are sometimes dispatched with no more than pointers to relevant
documentation topics.  Thus, what we say in this paper is in the
spirit of past ACL2 Workshop talks on ``What's New''.

We highlight documentation topics with the marker ``see :DOC''; for
example, see :DOC \href{http://www.cs.utexas.edu/users/moore/acl2/current/RELEASE-NOTES.html}{release-notes} and its subtopics (e.g., see :DOC
\href{http://www.cs.utexas.edu/users/moore/acl2/current/NOTE-3-6.html}{note-3-6} and see :DOC \href{http://www.cs.utexas.edu/users/moore/acl2/current/NOTE-4-3.html}{note-4-3} for changes introduced in ACL2 Versions
3.6 and 4.3, respectively).  We also refer to documentation topics
implicitly using underlining, for example: the topic {\tt
  \href{http://www.cs.utexas.edu/users/moore/acl2/current/ACL2-TUTORIAL.html}{\underline{acl2-tutorial}}}
is much improved.  For both kinds of references to documentation topics
(explicit and implicit), online copies of this paper have hyperlinks
to the documentation topic.

Those interested in implementation details are invited to see the
source code, which is extensively commented (see
Subsection~\ref{code-size}), available from the ACL2 home
page~\cite{acl2-home-page}.  In particular, each {\tt deflabel} form
for a release note has Lisp comments typically at a lower level than
the user documentation.

We present each enhancement by way of a question that we believe might
be asked by some ACL2 users, which is followed by an answer.  These
enhancements break naturally into categories.  We begin in
Section~\ref{programming}, which focuses on new programming features.
Next, Section~\ref{proofs} discusses enhancements pertaining to doing
proofs.  Finally, Section~\ref{system} addresses changes at the system
level.  We conclude with brief reflections.

\section{Programming Features}\label{programming}

In this section we describe several recent ACL2 enhancements that are
of particular use when programming.

\subsection{Equality variants}

\question{How can I avoid proving separate sets of rules for pairs of
  functions such as {\tt \href{http://www.cs.utexas.edu/users/moore/acl2/current/MEMBER.html}{\underline{member}}} and {\tt
    \href{http://www.cs.utexas.edu/users/moore/acl2/current/MEMBER-EQUAL.html}{\underline{member-equal}}}, which have logically equivalent
  definitions but use different equality tests in their definitions?}

To understand the question, recall that {\tt \href{http://www.cs.utexas.edu/users/moore/acl2/current/EQUAL.html}{\underline{equal}}}, {\tt \href{http://www.cs.utexas.edu/users/moore/acl2/current/EQL.html}{\underline{eql}}} and {\tt \href{http://www.cs.utexas.edu/users/moore/acl2/current/EQ.html}{\underline{eq}}} are logically
equivalent functions with different guards and different runtime efficiencies.  The user is expected to choose the variant that provides the most appropriate tradeoff between proof obligations and runtime performance.   The variants of
{\tt member} differ only by the equality test used; hence they may be proved equivalent but are not defined identically.  Now consider the following sequence
of two theorems, both proved automatically by ACL2.  Because the
function {\tt \href{http://www.cs.utexas.edu/users/moore/acl2/current/REVERSE.html}{\underline{reverse}}} is defined in terms of {\tt
  \href{http://www.cs.utexas.edu/users/moore/acl2/current/REVAPPEND.html}{\underline{revappend}}}, ACL2 can automatically apply the first
theorem (as a rewrite rule) to prove the second theorem (without
using induction).

\begin{acl2}
(defthm member-revappend
  (iff (member a (revappend x y))
       (or (member a x)
           (member a y)))
  :hints (("Goal" :induct (revappend x y))))

(defthm member-reverse
  (iff (member a (reverse x))
       (member a x)))
\end{acl2}

But the corresponding theorem about {\tt \href{http://www.cs.utexas.edu/users/moore/acl2/current/MEMBER-EQUAL.html}{\underline{member-equal}}},
just below, fails to be proved in ACL2 versions preceding 4.3, where
{\tt \href{http://www.cs.utexas.edu/users/moore/acl2/current/MEMBER.html}{\underline{member}}} and {\tt \href{http://www.cs.utexas.edu/users/moore/acl2/current/MEMBER-EQUAL.html}{\underline{member-equal}}} were
essentially different functions: their recursive definitions were
similar but differed in the equality test used ({\tt \href{http://www.cs.utexas.edu/users/moore/acl2/current/EQL.html}{\underline{eql}}}
or {\tt \href{http://www.cs.utexas.edu/users/moore/acl2/current/EQUAL.html}{\underline{equal}}}, respectively).

\begin{acl2}
(defthm member-equal-reverse
  (iff (member-equal a (reverse x))
       (member-equal a x)))
\end{acl2}

However, after Version 4.2, the proof succeeds for {\tt
  member\--equal\--reverse}.  Indeed, if {\tt member\--reverse} is proved
first, then it is applied in the proof of {\tt
  member\--equal\--reverse}.  The upshot is that we no longer need to
create separate libraries of rules for {\tt member} and {\tt
  member-equal}.

Briefly put, the change is that {\tt \href{http://www.cs.utexas.edu/users/moore/acl2/current/MEMBER.html}{\underline{member}}} is a macro
that generates a call of {\tt \href{http://www.cs.utexas.edu/users/moore/acl2/current/MEMBER-EQUAL.html}{\underline{member-equal}}} in the logic.
Here is a log showing in some detail the macroexpansion of calls of
{\tt \href{http://www.cs.utexas.edu/users/moore/acl2/current/MEMBER-EQ.html}{\underline{member-eq}}} and {\tt \href{http://www.cs.utexas.edu/users/moore/acl2/current/MEMBER.html}{\underline{member}}}, using {\tt
  :\href{http://www.cs.utexas.edu/users/moore/acl2/current/TRANS1.html}{\underline{trans1}}}.

\begin{acl2}
ACL2 !>:trans1 (member-eq a x)
 (MEMBER A X :TEST 'EQ)
ACL2 !>:trans1 (member a x :test 'eq)
 (LET-MBE ((X A) (L X))
          :LOGIC (MEMBER-EQUAL X L)
          :EXEC (MEMBER-EQ-EXEC X L))
ACL2 !>:trans1 (let-mbe ((x a) (l x))
                        :logic (member-equal x l)
                        :exec (member-eq-exec x l))
 (LET ((X A) (L X))
      (MBE :LOGIC (MEMBER-EQUAL X L)
           :EXEC (MEMBER-EQ-EXEC X L)))
ACL2 !>
\end{acl2}

As seen above, calls of macros {\tt \href{http://www.cs.utexas.edu/users/moore/acl2/current/MEMBER-EQ.html}{\underline{member-eq}}} and {\tt
  \href{http://www.cs.utexas.edu/users/moore/acl2/current/MEMBER.html}{\underline{member}}} ultimately generate calls of the function {\tt
  \href{http://www.cs.utexas.edu/users/moore/acl2/current/MEMBER-EQUAL.html}{\underline{member-equal}}} within the {\tt :logic} component of an
{\tt \href{http://www.cs.utexas.edu/users/moore/acl2/current/MBE.html}{\underline{mbe}}} call.  Many parts of the ACL2 reasoning engine
reduce an {\tt \href{http://www.cs.utexas.edu/users/moore/acl2/current/MBE.html}{\underline{mbe}}} call to its {\tt :logic} component; so
it is fair to say that the ACL2 prover treats a call of {\tt
  \href{http://www.cs.utexas.edu/users/moore/acl2/current/MEMBER-EQ.html}{\underline{member-eq}}} or {\tt \href{http://www.cs.utexas.edu/users/moore/acl2/current/MEMBER.html}{\underline{member}}} as a
corresponding call of the function {\tt \href{http://www.cs.utexas.edu/users/moore/acl2/current/MEMBER-EQUAL.html}{\underline{member-equal}}}.
Indeed, as part of this change we extended such reduction of {\tt
  \href{http://www.cs.utexas.edu/users/moore/acl2/current/MBE.html}{\underline{mbe}}} calls to additional contexts (for more on this,
search the documentation for references to ``guard holder'').

For more information about uniform treatment of functions whose
definitions differ only on the equality predicates used, including a
full listing of such functions, see :DOC \href{http://www.cs.utexas.edu/users/moore/acl2/current/EQUALITY-VARIANTS.html}{equality-variants}.

\subsection{\href{http://www.cs.utexas.edu/users/moore/acl2/current/DEFATTACH.html}{\underline{Defattach}}}

\question{How can I execute encapsulated functions, modify certain
  built-in function behavior, or program using refinements?}

The {\tt \href{http://www.cs.utexas.edu/users/moore/acl2/current/DEFATTACH.html}{\underline{defattach}}}
utility~\cite{defattach-workshop,defattach} provides all of the above.
Consider for example the following sequence of events, which
introduces a ``fold'' function that applies a given
associative-commutative function to successive members of a list.

\begin{acl2}
(encapsulate
 (((ac-fn * *) => * 
   :formals (x y)
   :guard (and (acl2-numberp x) 
               (acl2-numberp y))))
 (local (defun ac-fn (x y)
          (+ x y)))
 (defthm ac-fn-commutative
   (equal (ac-fn x y)
          (ac-fn y x)))
 (defthm ac-fn-associative
   (equal (ac-fn (ac-fn x y) z)
          (ac-fn x (ac-fn y z)))))

(defun fold (lst root)
  (cond ((endp lst) root)
        (t (fold (cdr lst)
                 (ac-fn (car lst) root)))))
\end{acl2}
At this point, evaluation of {\tt (fold '(2 3 4 5) 1)} fails, because
{\tt fold} calls {\tt ac-fn}, which is not defined.  But if we {\em
  attach} the built-in ACL2 multiplication function to {\tt ac-fn} we
can do such evaluation, as shown below.  Indeed, we can use evaluation
to explore conjectures, such as whether the value returned by a call
of {\tt fold} is unchanged if its first argument is reversed.  We omit
the output from  the call of {\tt defattach}, which shows proof
obligations being discharged.

\begin{acl2}
ACL2 !>(defattach ac-fn binary-*)
{\em{[[.. output omitted..]]}}
ACL2 !>(fold '(2 3 4 5) 1)
120
ACL2 !>(fold (reverse '(2 3 4 5)) 1)
120
ACL2 !>
\end{acl2}

Note however that attachments are not invoked during proofs.
Continuing with the example above, the proof fails for {\tt (thm
  (equal (fold '(2 3 4 5) 1) 120))}.  Indeed, because attachments can
be overwritten with new attachments it is important that they are
turned off not only for proofs but also for other logical contexts,
such as the evaluation of {\tt defconst} forms.

The discussion above shows how {\tt defattach} supports execution of
encapsulated function calls and gives a hint about refinement.  But a
third use is the modification of built-in function behavior, towards
opening up the architecture of ACL2.  Certain ACL2 prover functions
are now implemented with {\tt defattach} (see source file {\tt
  boot\--strap\--pass\--2.lisp}), permitting the user to customize some
heuristics by attaching other functions to them.  We invite the user community to
request more such support.  One example is the built-in function {\tt
  ancestors\--check}, which implements a rewriting heuristic.  Robert
Krug requested that this function be attachable, and we thank him for
that; actually he went further and provided the necessary proof
support.

There is much more to know about {\tt defattach}, but our goal in this
paper is simply to provide an introduction to it.  To learn more see
:DOC \href{http://www.cs.utexas.edu/users/moore/acl2/current/DEFATTACH.html}{defattach}.  For logical foundations and (significant)
implementation subtleties, see a comment in the ACL2 source code
entitled ``Essay on Defattach'', which explains the subtle logical
foundations of {\tt defattach}, and will ultimately be incorporated
into a comprehensive treatment~\cite{defattach}.\footnote{We thank
  C\'esar Mu\~noz and Shankar for useful email discussions on relationships between
  the {\tt defattach} feature of ACL2 and the PVS features of {\tt
    defattach}, theory interpretations, and theory parameters with
  assumptions.  We expect to explore these relationships in the
  aforementioned paper.}

\subsection{\href{http://www.cs.utexas.edu/users/moore/acl2/current/RETURN-LAST.html}{\underline{Return-last}}}

\question{How can I arrange that my macros have raw-Lisp side
  effects, like time\$?}

Recall that {\tt (time\$ form)} is semantically the same as {\tt
  form}, except that timing information is printed to the terminal
after evaluation is complete.  To see how this works, we consider the
macroexpansion of a call of {\tt
  \href{http://www.cs.utexas.edu/users/moore/acl2/current/TIME$.html}{\underline{time\$}}}.
Note: the interpretation of {\tt (list 0 nil nil nil nil)} is not
important for this explanation.

\begin{acl2}
ACL2 !>:trans1 (time\$ (foo 3 4))
 (TIME\$1 (LIST 0 NIL NIL NIL NIL)
         (FOO 3 4))
ACL2 !>:trans1 (time\$1 (list 0 nil nil nil nil)
                       (foo 3 4))
 (RETURN-LAST 'TIME\$1-RAW
              (LIST 0 NIL NIL NIL NIL)
              (FOO 3 4))
ACL2 !>
\end{acl2}

In the logic, {\tt \href{http://www.cs.utexas.edu/users/moore/acl2/current/RETURN-LAST.html}{\underline{return-last}}} is a function that returns
its last argument (as its name suggests).  But in raw Lisp, {\tt
  return-last} is a macro.  In essence, it expands to a call of the
(unquoted) first argument on the remaining two arguments, which should
return the value of the last argument.

\begin{acl2}
ACL2 !>:q

Exiting the ACL2 read-eval-print loop.  To re-enter, execute (LP).
? [RAW LISP] (macroexpand-1
              '(return-last 'time\$1-raw
                            (list 0 nil nil nil nil)
                            (foo 3 4)))
(TIME\$1-RAW (LIST 0 NIL NIL NIL NIL) (FOO 3 4))
T
? [RAW LISP] 
\end{acl2}

The raw Lisp macro {\tt time\$1-raw} is what actually carries out the
timing of the indicated call of {\tt foo} above.

Note that there must be an active trust tag (see :DOC
\href{http://www.cs.utexas.edu/users/moore/acl2/current/DEFTTAG.html}{defttag})
in order to extend the special treatment of {\tt return-last} to
additional values of its first argument.  It is the user's
responsibility, when making such an extension, to ensure that any call
of the value of the first argument does indeed return the value of the
last argument --- which brings us back to the original question,
above, which is how to make such an extension.

A macro {\tt
  \href{http://www.cs.utexas.edu/users/moore/acl2/current/DEFMACRO-LAST.html}{\underline{defmacro-last}}}
makes it a rather simple undertaking to make such an extension.  The
distributed book {\tt books/\-misc/\-profiling.lisp} illustrates how
this works by defining an ACL2 macro, {\tt with-profiling}, together
with a raw-Lisp macro {\tt with-profiling-raw} that causes the desired
side effects.  For more explanation of this example, and of {\tt
  defmacro-last} and {\tt return-last}, see :DOC
\href{http://www.cs.utexas.edu/users/moore/acl2/current/RETURN-LAST.html}{return-last}.

\begin{acl2}
{\em{; A trust tag is needed for {\href{http://www.cs.utexas.edu/users/moore/acl2/current/PROGN!.html}{\underline{progn!}}}; see :DOC \href{http://www.cs.utexas.edu/users/moore/acl2/current/DEFTTAG.html}{defttag}.}}
(defttag :profiling)

(progn!
 (set-raw-mode t)
 (load (concatenate 'string 
                    (cbd)
                    "profiling-raw.lsp")))

(defmacro-last with-profiling)
\end{acl2}

Additional examples show the flexibility of {\tt defmacro-last}.  Sol
Swords and Jared Davis have used {\tt defmacro-last} to create a macro
{\tt with-fast\--alist} for the HONS version of ACL2, which is defined
and documented in the book {\tt centaur/\-misc/\-hons\--extra.lisp}
distributed in the acl2-books svn repository~\cite{acl2-books-svn}.
David Rager has used {\tt defmacro-last} to create a timing utility
that shows garbage collection information, distributed with ACL2 as
{\tt books/\-tools/\-time\--dollar\--with\--gc.lisp}.  Both books, as
well as the book {\tt profiling.lisp} mentioned above, come with
associated raw Lisp files that implement the desired side effects.

\subsection{Avoiding guard violations}

\question{How can I avoid errors on ill-guarded calls, even in raw
  Lisp, and even for {\tt :\href{http://www.cs.utexas.edu/users/moore/acl2/current/PROGRAM.html}{\underline{program}}} mode functions?}

See :DOC \href{http://www.cs.utexas.edu/users/moore/acl2/current/WITH-GUARD-CHECKING.html}{with-guard-checking} and see :DOC \href{http://www.cs.utexas.edu/users/moore/acl2/current/EC-CALL.html}{ec-call}.  The latter was
introduced in ACL2 Version 3.4, and replaces a call with its so-called
``executable-counterpart''.  But {\tt with\--guard\--checking} is
newer (introduced in Version 4.0), and can be used to suppress guard
checking for executable counterparts.  The following example
illustrates how these two work together to answer the above question.

\begin{acl2}
ACL2 !>(defun foo (x)
         (declare (xargs :mode :program))
         (with-guard-checking nil (ec-call (car x))))

Summary
Form:  ( DEFUN FOO ...)
Rules: NIL
Time:  0.00 seconds (prove: 0.00, print: 0.00, other: 0.00)
 FOO
ACL2 !>(foo 3)
NIL
ACL2 !>
\end{acl2}

Note that the use of {\tt \href{http://www.cs.utexas.edu/users/moore/acl2/current/EC-CALL.html}{\underline{ec-call}}} is necessary in order
to avoid calling {\tt car} on {\tt 3} in raw Lisp.  If instead {\tt foo}
were defined in {\tt :\href{http://www.cs.utexas.edu/users/moore/acl2/current/LOGIC.html}{\underline{logic}}} mode, then the use of {\tt
  \href{http://www.cs.utexas.edu/users/moore/acl2/current/EC-CALL.html}{\underline{ec-call}}} would not be necessary above because the
executable-counterpart of {\tt car} would be called on {\tt 3}.

For background about how guards and evaluation work, see :DOC \href{http://www.cs.utexas.edu/users/moore/acl2/current/GUARD.html}{guard}
and its subtopics; in particular see :DOC \href{http://www.cs.utexas.edu/users/moore/acl2/current/GUARDS-AND-EVALUATION.html}{guards-and-evaluation} and
see :DOC \href{http://www.cs.utexas.edu/users/moore/acl2/current/GUARD-EVALUATION-TABLE.html}{guard-evaluation-table}.

\subsection{Printing without state}

\question{I'm doing printing without accessing state.  How can I avoid
  producing messages when proof output is turned off?  More generally,
  how best can I print without actually reading or writing the state?}

Macros {\tt \href{http://www.cs.utexas.edu/users/moore/acl2/current/OBSERVATION-CW.html}{\underline{observation-cw}}} and {\tt warning\$-cw} are
analogues of macros {\tt \href{http://www.cs.utexas.edu/users/moore/acl2/current/OBSERVATION.html}{\underline{observation}}} and {\tt warning\$}
that, however, do not access {\tt \href{http://www.cs.utexas.edu/users/moore/acl2/current/STATE.html}{\underline{state}}}.  We strongly
suggest using these in place of the macro {\tt \href{http://www.cs.utexas.edu/users/moore/acl2/current/CW.html}{\underline{cw}}} in
functions called during a proof, for example during evaluation of
clause-processors or computed hints, so that users can turn off such
messages by using {\tt \href{http://www.cs.utexas.edu/users/moore/acl2/current/SET-INHIBIT-OUTPUT-LST.html}{\underline{set-inhibit-output-lst}}}.

{\em Remarks.}  (1) The above two macros are implemented using
{\href{http://www.cs.utexas.edu/users/moore/acl2/current/WORMHOLE.html}{\underline{wormhole}}}s, which were given an improved implementation
in Version 4.0 and later.  (2) There are now many utilities with the
suffix ``{\tt{-cmp}}''.  These traffic in so-called ``context-message
pairs'', rather than {\href{http://www.cs.utexas.edu/users/moore/acl2/current/STATE.html}{\underline{state}}}, as described in the ``Essay
on Context-message Pairs'' in the ACL2 source code (file {\tt
  translate.lisp}).

\question{How can I create a string using formatted printing
  functions, without printing out the string and preferably without
  accessing the ACL2 state?}

See :DOC \href{http://www.cs.utexas.edu/users/moore/acl2/current/PRINTING-TO-STRINGS.html}{printing-to-strings} for analogues of functions like {\tt
  \href{http://www.cs.utexas.edu/users/moore/acl2/current/FMT.html}{\underline{fmt}}} that return strings and do not access state.  For
example:

\begin{acl2}
ACL2 !>(fmt1-to-string \verb|"Hello, ~x0"|
                       \verb|(list (cons #\0 'world))|
                       \verb|0|)
\verb|(12 "Hello, WORLD")|
ACL2 !>
\end{acl2}

Also see :DOC \href{http://www.cs.utexas.edu/users/moore/acl2/current/IO.html}{io} for a discussion of how to open a channel that
connects to a string, along with an associated utility for retrieving
the string printed to that channel, {\tt
  get-\-output\--stream\--string\$} (which however does access the
ACL2 {\href{http://www.cs.utexas.edu/users/moore/acl2/current/STATE.html}{\underline{state}}}).

{\em Remark for system developers.}  If you are willing to use trust
tags (see :DOC \href{http://www.cs.utexas.edu/users/moore/acl2/current/DEFTTAG.html}{defttag}), then see :DOC \href{http://www.cs.utexas.edu/users/moore/acl2/current/WITH-LOCAL-STATE.html}{with-local-state} for a
potentially unsound utility that allows you to create a temporary ACL2
state object out of thin air!

\subsection{Parallel evaluation}

\question{How can I build an application that evaluates code in
  parallel?}

ACL2(p) is an experimental extension of ACL2 that incorporates
research and code from David
Rager~\cite{acl2-par-workshop,09-rager-hunt}.  Recent additions
include a macro {\tt \href{http://www.cs.utexas.edu/users/moore/acl2/current/SPEC-MV-LET.html}{\underline{spec-mv-let}}}, which allows
speculative evaluation in parallel.  See :DOC \href{http://www.cs.utexas.edu/users/moore/acl2/current/PARALLELISM.html}{parallelism}.  Later
below we discuss parallel proofs of subgoals.

\subsection{Other recent programming support}

\question{How can I get around some syntactic restrictions imposed by the
use of multiple values?}

The macros {\tt \href{http://www.cs.utexas.edu/users/moore/acl2/current/MV?.html}{\underline{mv?}}} and {\tt \href{http://www.cs.utexas.edu/users/moore/acl2/current/MV-LET?.html}{\underline{mv-let?}}} are
analogues of {\tt \href{http://www.cs.utexas.edu/users/moore/acl2/current/MV.html}{\underline{mv}}} and {\tt \href{http://www.cs.utexas.edu/users/moore/acl2/current/MV-LET.html}{\underline{mv-let}}} which,
however, may return or bind just one variable (respectively).

The function {\tt mv-list} converts multiple values to a single value
that is a list, for example as follows.
\begin{acl2}
ACL2 !>(mv-list 3 (mv 5 6 7))
(5 6 7)
ACL2 !>(cdr (mv-list 3 (mv 5 6 7)))
(6 7)
ACL2 !>
\end{acl2}

\question{How can I redefine system functions and macros inside the ACL2 loop?}

A utility for this purpose, {\tt redef+}, is now an embedded event
form (i.e., it can go in {\href{http://www.cs.utexas.edu/users/moore/acl2/current/BOOKS.html}{\underline{books}}}).  Note that the
counterpart of {\tt :redef+}, {\tt :redef-}, now turns off
redefinition (it formerly had not done so).

\question{What support is provided for tracing function calls inside
  the ACL2 loop?}

See :DOC \href{http://www.cs.utexas.edu/users/moore/acl2/current/TRACE$.html}{trace\$}.  Although this utility has been
around for many years, it has benefited from recent improvements.

\question{What other recent ACL2 programming enhancements might I be
  missing?}

See :DOC \href{http://www.cs.utexas.edu/users/moore/acl2/current/RELEASE-NOTES.html}{release-notes}.  Useful new features include the following.

\begin{itemize}

\item The macro {\tt \href{http://www.cs.utexas.edu/users/moore/acl2/current/TIME$.html}{\underline{time\$}}} is now user-customizable
  (thanks to an initial implementation contributed by Jared Davis).

\item The function {\tt \href{http://www.cs.utexas.edu/users/moore/acl2/current/PKG-IMPORTS.html}{\underline{pkg-imports}}} returns the list of
  symbols imported into a specified package.

\item {\tt (File-write-date\$ filename state)} returns the Common Lisp
  file-write-date of the given filename.

\item The macro {\tt \href{http://www.cs.utexas.edu/users/moore/acl2/current/APPEND.html}{\underline{append}}} no longer requires two or
  more arguments: now {\tt (append)} expands to {\tt nil}, and {\tt
    (append X)} expands to {\tt X}.

\end{itemize}

\section{Proof Debug, Control, and \\ Reporting}\label{proofs}

This section addresses recent ACL2 improvements in user interaction
with the ACL2 prover.

\subsection{Hints}

The {\href{http://www.cs.utexas.edu/users/moore/acl2/current/HINTS.html}{\underline{hints}}} mechanism continues to become more flexible
and better documented.  In Subsection~\ref{proof-checker} we discuss
one major improvement, the use of the {\tt :instructions} keyword in
hints; but first we point out several other advances in hints.

The first two new features mentioned below, override-hints and
backtrack hints, have been used to integrate testing with the ACL2
prover~\cite{acl2-testing}.

\question{How can I provide default hints that are not ignored when I
  give explicit hints to goals?}

See :DOC \href{http://www.cs.utexas.edu/users/moore/acl2/current/OVERRIDE-HINTS.html}{override-hints}.

\question{The hints mechanism has always confused me a bit; for
  example, some hints are inherited by subgoals and others are not.
  How can I better understand the ``flow'' of hints?}

See :DOC \href{http://www.cs.utexas.edu/users/moore/acl2/current/HINTS-AND-THE-WATERFALL.html}{hints-and-the-waterfall} for a detailed explanation of how
hints are processed.  Also, some helpful examples may be found in
distributed book {\tt books/\-hints/\-basic\--tests.lisp}.

\question{How can I write a computed hint that can backtrack if
  `undesirable' subgoals are created?}

See :DOC
\href{http://www.cs.utexas.edu/users/moore/acl2/current/HINTS.html}{hints}
for a discussion of {\tt :backtrack} hints.

\question{How can I program up fancy computed hints that do not keep
  announcing ``thanks'' each time one is applied?}

See :DOC \href{http://www.cs.utexas.edu/users/moore/acl2/current/HINTS.html}{hints} for a discussion of {\tt :no-thanks} hints.

\question{I know how to limit backtracking in the rewriter by using
  {\tt \href{http://www.cs.utexas.edu/users/moore/acl2/current/SET-BACKCHAIN-LIMIT.html}{\underline{set-backchain-limit}}}, but how can I do this at the
  level of hints?}

See :DOC \href{http://www.cs.utexas.edu/users/moore/acl2/current/HINTS.html}{hints} for a discussion of {\tt :backchain-limit-rw}.

\subsection{Proof-checker enhancements}\label{proof-checker}

\question{How can I better employ the {\href{http://www.cs.utexas.edu/users/moore/acl2/current/PROOF-CHECKER.html}{\underline{proof-checker}}} to
  create proper ACL2 {\href{http://www.cs.utexas.edu/users/moore/acl2/current/EVENTS.html}{\underline{events}}}?}

See :DOC \href{http://www.cs.utexas.edu/users/moore/acl2/current/PROOF-CHECKER.html}{proof-checker} for a (long-standing) utility for conducting
proofs interactively.  Probably the most common use of the
proof-checker is to invoke {\tt \href{http://www.cs.utexas.edu/users/moore/acl2/current/VERIFY.html}{\underline{verify}}} to explore the
proof of a conjecture whose automated attempt has failed.  But
sometimes it is convenient to save a proof-checker proof using its
{\tt :exit} command, creating an event by pasting that proof as the
value of an {\tt :\href{http://www.cs.utexas.edu/users/moore/acl2/current/INSTRUCTIONS.html}{\underline{instructions}}} keyword.  An example is
given below.

The proof-checker's use in the creation of {\href{http://www.cs.utexas.edu/users/moore/acl2/current/EVENTS.html}{\underline{events}}} has
recently been made more flexible in two ways.

\begin{itemize}

\item User-defined macro commands (see :DOC \href{http://www.cs.utexas.edu/users/moore/acl2/current/DEFINE-PC-MACRO.html}{define-pc-macro}) are now
  legal for {\tt :{\href{http://www.cs.utexas.edu/users/moore/acl2/current/INSTRUCTIONS.html}{\underline{instructions}}}}.

\item The use of an {\tt :instructions} keyword is now supported inside
  {\tt :hints}, in particular at the subgoal level.

\end{itemize}

Below is an example, inspired by the event {\tt
  not\--equal\--intern\--in\--package\--of\--symbol\--nil} from the
distributed book {\tt books/\-coi/\-gensym/\-gensym.lisp}.  You'll see
that we exit the proof-checker when we've gotten past the sticky bit,
and that we use the new capability for putting {\tt :instructions}
inside {\tt :hints} (though that's not actually needed for this
example).

\begin{acl2}
ACL2 !>(verify
        (implies
         (and (stringp string)
              (symbolp symbol)
              (equal (intern-in-package-of-symbol string symbol)
                     nil))
         (equal string "NIL")))
->: bash
{\em{[[.. output omitted; simplified to one goal ..]]}}
->: th {\em ; show current goal's hypotheses and conclusion}
*** Top-level hypotheses:
1. (STRINGP STRING)
2. (SYMBOLP SYMBOL)
3. (NOT (INTERN-IN-PACKAGE-OF-SYMBOL STRING SYMBOL))

The current subterm is:
(EQUAL STRING "NIL")
->: (casesplit {\em ; split into two goals, by cases}
     (not {\em ; using the negation makes example more interesting}
      (equal (symbol-name
              (intern-in-package-of-symbol string symbol))
             string)))

Creating one new goal:  ((MAIN . 1) . 1).
->: prove
{\em{[[.. output omitted; the proof fails ..]]}}
->: th
*** Top-level hypotheses:
1. (STRINGP STRING)
2. (SYMBOLP SYMBOL)
3. (NOT (INTERN-IN-PACKAGE-OF-SYMBOL STRING SYMBOL))
4. (NOT (EQUAL (SYMBOL-NAME (INTERN-IN-PACKAGE-OF-SYMBOL STRING
                                                         SYMBOL))
               STRING))

The current subterm is:
(EQUAL STRING "NIL")
->: (drop 3) {\em ; Drop the third hypothesis.}
{\em{; Hypothesis 4 is false, but hypothesis 3 gets in the way.}}
->: prove
***** Now entering the theorem prover *****

But simplification reduces this to T, using primitive type
reasoning and the :rewrite rule
SYMBOL-NAME-INTERN-IN-PACKAGE-OF-SYMBOL.

Q.E.D.

The proof of the current goal, (MAIN . 1), has been completed.
However, the following subgoals remain to be proved:
  ((MAIN . 1) . 1).
Now proving ((MAIN . 1) . 1).
->: (exit t)

Not exiting, as there remain unproved goals:  ((MAIN . 1) . 1).
The original goal is:
    (IMPLIES (AND (STRINGP STRING)
                  (SYMBOLP SYMBOL)
                  (EQUAL (INTERN-IN-PACKAGE-OF-SYMBOL STRING
                                                      SYMBOL)
                         NIL))
             (EQUAL STRING "NIL"))
  Here is the current instruction list, starting with the first:
    (:BASH
     (:CASESPLIT
        (NOT (EQUAL (SYMBOL-NAME 
                     (INTERN-IN-PACKAGE-OF-SYMBOL STRING SYMBOL))
                    STRING)))
     (:DROP 3)
     :PROVE)
->: exit
Exiting....
 NIL
{\em{; Now we can paste in the above instructions and prove the theorem.}}
ACL2 !>(thm
        (implies
         (and (stringp string)
              (symbolp symbol)
              (equal (intern-in-package-of-symbol string symbol)
                     nil))
         (equal string "NIL"))
        :hints
        (("Goal"
          :instructions
          (:BASH
            (:CASESPLIT
              (NOT (EQUAL (SYMBOL-NAME (INTERN-IN-PACKAGE-OF-SYMBOL
                                        STRING SYMBOL))
                          STRING)))
            (:DROP 3)
            :PROVE))))

[Note:  A hint was supplied for our processing of the goal above. 
Thanks!]

We now apply the trusted :CLAUSE-PROCESSOR function
PROOF-CHECKER-CL-PROC to produce one new subgoal.

Goal'
(IMPLIES
     (AND (STRINGP STRING)
          (SYMBOLP SYMBOL)
          (NOT (INTERN-IN-PACKAGE-OF-SYMBOL STRING SYMBOL))
          (EQUAL (SYMBOL-NAME (INTERN-IN-PACKAGE-OF-SYMBOL STRING
                                                           SYMBOL))
                 STRING))
     (EQUAL STRING "NIL")).

But simplification reduces this to T, using the
:executable-counterpart of SYMBOL-NAME.

Q.E.D.

Summary
Form:  ( THM ...)
Rules: ((:EXECUTABLE-COUNTERPART SYMBOL-NAME))
Time:  0.00 seconds (prove: 0.00, print: 0.00, other: 0.00)
Prover steps counted:  69

Proof succeeded.
ACL2 !>
\end{acl2}

\subsection{Parallelism in proofs}

\question{Can I speed up proofs by having subgoals proved in
  parallel?}

Yes, if you build the experimental extension for parallelism that
incorporates David Rager's dissertation work~\cite{rager-diss}.  See
:DOC \href{http://www.cs.utexas.edu/users/moore/acl2/current/PARALLELISM.html}{parallelism}.

\subsection{Limiting proof effort}

\question{I like using {\tt with-prover-time-limit} to limit proof
  effort, but is there something similar that is
  platform-independent?}

See :DOC \href{http://www.cs.utexas.edu/users/moore/acl2/current/WITH-PROVER-STEP-LIMIT.html}{with-prover-step-limit}.  Also see :DOC \href{http://www.cs.utexas.edu/users/moore/acl2/current/SET-PROVER-STEP-LIMIT.html}{set-prover-step-limit},
which lets you set the default limit for the current environment
(whether it be at the top level, or in an {\tt encapsulate}, {\tt
  progn}, {\tt make-event}, {\tt certify-book}, etc.).

Note that {\tt \href{http://www.cs.utexas.edu/users/moore/acl2/current/WITH-PROVER-STEP-LIMIT.html}{\underline{with-prover-step-limit}}} may be used to form
{\href{http://www.cs.utexas.edu/users/moore/acl2/current/EVENTS.html}{\underline{events}}} in {\href{http://www.cs.utexas.edu/users/moore/acl2/current/BOOKS.html}{\underline{books}}}.  The same is now true
(but had not been in the past) for {\tt
  \href{http://www.cs.utexas.edu/users/moore/acl2/current/WITH-PROVER-TIME-LIMIT.html}{\underline{with-prover-time-limit}}}.

\subsection{Proof debugging}

\question{I formerly used {\tt \href{http://www.cs.utexas.edu/users/moore/acl2/current/ACCUMULATED-PERSISTENCE.html}{\underline{accumulated-persistence}}},
  but its output seemed too limited.  Are there any new options that
  could make it more useful?}

By default, {\tt \href{http://www.cs.utexas.edu/users/moore/acl2/current/SHOW-ACCUMULATED-PERSISTENCE.html}{\underline{show-accumulated-persistence}}} now breaks
down the statistics by ``useful'' and ``useless'' applications of the
rules.  If you enable the feature with {\tt (accumulated\--persistence
  :all)}, then statistics are further broken down by rule hypothesis
and conclusion.

Also, see :DOC \href{http://www.cs.utexas.edu/users/moore/acl2/current/ACCUMULATED-PERSISTENCE.html}{accumulated-persistence} for a discussion of the {\tt
  :runes} option for obtaining a raw, alphabetical listing.

\question{A {\tt defthm} failed in the middle of an {\tt encapsulate}
  or {\tt certify-book}.  How can I get into a state where I can work
  on the failed proof?}

{\tt \href{http://www.cs.utexas.edu/users/moore/acl2/current/REDO-FLAT.html}{\underline{Redo-flat}}} has been around since Version 3.0.1, but
among the latest improvements is that now it works for {\tt
  certify-book}.

\question{How can I get debug-level information on what's going on
  with forward-chaining?}

See :DOC \href{http://www.cs.utexas.edu/users/moore/acl2/current/FORWARD-CHAINING-REPORTS.html}{forward-chaining-reports}.

\question{How do I control all the noise I get from proofs?}

Starting with Version 4.0, you can inhibit specified parts of the
Summary printed at the conclusion of an event; see :DOC
\href{http://www.cs.utexas.edu/users/moore/acl2/current/SET-INHIBITED-SUMMARY-TYPES.html}{set-inhibited-summary-types}.
For example, ACL2 developers sometimes evaluate the form
\begin{acl2p}
(set-inhibited-summary-types '(time))
\end{acl2p}
to compare proof output from two runs without the distraction of time
differences.

But the most important recent such development is a bit older,
introduced in Version 3.3: {\href{http://www.cs.utexas.edu/users/moore/acl2/current/GAG-MODE.html}{\underline{gag-mode}}}.  Gag-mode allows
you to turn off all but key prover output, so that you can focus on
key checkpoints (see :DOC \href{http://www.cs.utexas.edu/users/moore/acl2/current/THE-METHOD.html}{the-method} and see :DOC
\href{http://www.cs.utexas.edu/users/moore/acl2/current/INTRODUCTION-TO-THE-THEOREM-PROVER.html}{introduction\--to\--the\--theorem\--prover}).  We may well make
gag-mode the default at some point in the future.

Two improvements to gag-mode were introduced with Version 4.3.  (1)
The printing of induction schemes is suppressed in gag-mode.  (2) You
can now limit the printing of subgoal names when using {\tt
  :set-gag-mode :goals}; see :DOC
\href{http://www.cs.utexas.edu/users/moore/acl2/current/SET-PRINT-CLAUSE-IDS.html}{set-print-clause-ids}.

\subsection{New heuristics}

\question{Are there any new developments in proof heuristics?}

There are two significant new heuristics in Version 4.3.

ACL2 now caches information for failed applications of
{\href{http://www.cs.utexas.edu/users/moore/acl2/current/REWRITE.html}{\underline{rewrite}}}
rules.  We have seen a speedup of 11\% on the ACL2 regression suite,
but in some cases the speedup is significantly higher.  See :DOC
\href{http://www.cs.utexas.edu/users/moore/acl2/current/SET-RW-CACHE-STATE.html}{set-rw-cache-state}
for information about controlling this feature, including information
on how to turn it off in the very unlikely case that it makes a proof
fail.

Our description of the second heuristic relies on an understanding of
free variables in hypotheses of rules; see :DOC
\href{http://www.cs.utexas.edu/users/moore/acl2/current/FREE-VARIABLES.html}{free-variables}.
Since Version 2.2 (November, 2002), ACL2 has by default considered
every match from the current context for free variables in a
hypothesis of a {\underline{rewrite}}, {\underline{linear}}, or
{\underline{forward-chaining}} rule, until finding a match for which
the rule's hypotheses are all discharged.  Now, that behavior is also
the default for \underline{type-prescription} rules; see :DOC
\href{http://www.cs.utexas.edu/users/moore/acl2/current/FREE-VARIABLES-TYPE-PRESCRIPTION.html}{free-variables-type-prescription}.

\question{I'd like to learn more about how to use the ACL2 prover
  effectively, and I'm willing to do some reading about that.  But
  where should I start?}

The
{\href{http://www.cs.utexas.edu/users/moore/acl2/current/ACL2-TUTORIAL.html}{\underline{acl2-tutorial}}}
:DOC topic has been significantly expanded and improved.  It contains
pointers to different materials that you may choose to read, depending
on your learning style.  In particular, see :DOC
\href{http://www.cs.utexas.edu/users/moore/acl2/current/INTRODUCTION-TO-THE-THEOREM-PROVER.html}{introduction-to-the-theorem-prover}
for a tutorial on how to use the ACL2 prover effectively.

\section{System-level Enhancements}\label{system}

Here we discuss a few infrastructural improvements other than direct
support of programming and proofs.  Most experienced ACL2 users
consider system infrastructure an important component of ACL2's
usability.

\subsection{Two-run certification to avoid trust tags}

The first question below is only likely to be asked by system
builders.

\question{How can I certify a book that uses unverified proof tools whose solutions I know how
to check --- without making the book depend on a trust tag?}

See :DOC \href{http://www.cs.utexas.edu/users/moore/acl2/current/SET-WRITE-ACL2X.html}{set-write-acl2x}.

\subsection{Certifying a subset of the distributed books}

\question{How can I better control book certification?  In particular,
  I'd like to avoid certifying all the distributed books, since I only
  intend to include some of them.}

See :DOC \href{http://www.cs.utexas.edu/users/moore/acl2/current/BOOK-MAKEFILES.html}{book-makefiles} for answers to such infrastructural questions.
In particular, see the discussion there of environment variable {\tt
  ACL2\_BOOK\_DIRS}.

It has been the case for some time that by default, no
{\href{http://www.cs.utexas.edu/users/moore/acl2/current/ACL2-CUSTOMIZATION.html}{\underline{acl2-customization}}} file is loaded during `{\tt make
  regression}'.  The above documentation topic also mentions the new name for
an environment variable, now {\tt ACL2\_CUSTOMIZATION}, and explains
how it can be used to override that default behavior.

\subsection{Size and breakdown of ACL2 source code}\label{code-size}

\question{How big is the ACL2 source code?}

We now distribute a file {\tt doc/acl2-code-size.txt}.  Feel free to
poke around in the {\tt doc/} directory, or email the authors of ACL2,
if you want to use the same tools we use to compute size.  As of this
writing, here are the contents of the above file.

\begin{acl2}

CODE LINES:
   97465 lines,   4270000 characters
COMMENT LINES:
   51917 lines,   3062889 characters
BLANK LINES (excluding documentation):
   22884 lines,     24404 characters
DOCUMENTATION LINES:
   79543 lines,   3550075 characters
TOTAL:
  251809 lines,  10907368 characters
\end{acl2}

\subsection{An mbe restriction lifted}

\question{Is there any way to call {\tt \href{http://www.cs.utexas.edu/users/moore/acl2/current/MBE.html}{\underline{mbe}}} in the body
  of a definition within an {\tt \href{http://www.cs.utexas.edu/users/moore/acl2/current/ENCAPSULATE.html}{\underline{encapsulate}}} that has a
  non-empty {\href{http://www.cs.utexas.edu/users/moore/acl2/current/SIGNATURE.html}{\underline{signature}}}?}

Yes.  Some such restriction is necessary (see :DOC \href{http://www.cs.utexas.edu/users/moore/acl2/current/NOTE-3-4.html}{note-3-4}).
However, this restriction is now lifted provided you declare the
definition to be non-executable (typically by using {\tt \href{http://www.cs.utexas.edu/users/moore/acl2/current/DEFUN-NX.html}{\underline{defun-nx}}}).

\subsection{Aborting just one ld level}

\question{How can I avoid popping all the way back to the top level
  when I merely want to exit the {\tt :{\href{http://www.cs.utexas.edu/users/moore/acl2/current/BRR.html}{\underline{brr}}}}
  ``break-rewrite'' loop?}

Use {\tt :\href{http://www.cs.utexas.edu/users/moore/acl2/current/P!.html}{\underline{p!}}} instead of {\tt :\href{http://www.cs.utexas.edu/users/moore/acl2/current/A!.html}{\underline{a!}}}.  This same
trick works if you are in a nested call of {\tt {\href{http://www.cs.utexas.edu/users/moore/acl2/current/LD.html}{\underline{ld}}}}.

\section{Concluding remarks}

We believe that one of ACL2's greatest
strengths is its integration of programming and proof --- with due regard for both efficiency and soundness.
The ACL2 system continues to evolve through feedback from the ACL2
user community.  Many of the enhancements discussed here came about in
response to such feedback; see :DOC \href{http://www.cs.utexas.edu/users/moore/acl2/current/RELEASE-NOTES.html}{release-notes} to find specific
individuals associated with enhancement requests.  We very much
appreciate the opportunity to improve ACL2 in useful ways, and thus we
strongly encourage ACL2 users to let us know how we can make the
system more effective for them.

\section*{Acknowledgements}

We are grateful to the ACL2 user community for being primary drivers
of the development of ACL2.  We also thank Sandip Ray for helpful
feedback on a draft of this paper.

This material is based upon work supported the National Science
Foundation under Grant Nos. CCF-0945316 and CNS-0910913.

\bibliographystyle{eptcs}


\end{document}